\documentclass[11pt]{article}

\usepackage[utf8]{inputenc}
\usepackage[T1]{fontenc}
\usepackage{lmodern}
\usepackage[a4paper,margin=1in]{geometry}
\usepackage{amsmath,amssymb,amsfonts}
\usepackage{bm}
\usepackage{graphicx}
\usepackage{xcolor}
\usepackage{url}
\usepackage[colorlinks=true, allcolors=blue]{hyperref}
\usepackage{apacite} 
\usepackage{setspace}

\title{Significant Amplification of Turbulent Energy Dissipation through the Shock Transition at Mars}

\usepackage{authblk}

\author[1]{Wence Jiang}
\author[1,2]{Hui Li}
\author[3,4]{Nahuel Andr\'es}
\author[5]{Lina Hadid}
\author[6]{Daniel Verscharen}
\author[1,2,7]{Chi Wang}

\affil[1]{\small State Key Laboratory of Solar Activity and Space Weather, National Space Science Center, Chinese Academy of Sciences, Beijing 100190, China}
\affil[2]{\small School of Astronomy and Space Science, University of Chinese Academy of Sciences, Beijing 100190, China}
\affil[3]{\small Universidad de Buenos Aires, Facultad de Ciencias Exactas y Naturales, Departamento de F\'{i}sica, Buenos Aires, Argentina}
\affil[4]{\small CONICET - Universidad de Buenos Aires, Instituto de F\'isica Interdisciplinaria y Aplicada (INFINA), Buenos Aires, Argentina}
\affil[5]{\small Laboratoire de Physique des Plasmas, \'Ecole Polytechnique, CNRS, Sorbonne University, Observatoire de Paris, Univ. Paris--Sud, Palaiseau Cedex F-91128, France}
\affil[6]{\small Mullard Space Science Laboratory, University College London, Dorking RH5 6NT, UK}
\affil[7]{\small College of Earth and Planetary Sciences, University of Chinese Academy of Sciences, Beijing 100190, China}

\affil[ ]{\small Corresponding authors: \href{mailto:hli@nssc.ac.cn}{hli@nssc.ac.cn}, \href{mailto:jiangwence@swl.ac.cn}{jiangwence@swl.ac.cn}}

\date{} 

\begin{document}
\maketitle

\begin{abstract}
Turbulence is fundamental to energy transfer across scales in space and astrophysical plasmas. Bow shock interactions have long been hypothesized to significantly modify turbulence in planetary environments, yet the quantification of such effects and their parametric dependencies remain largely unaddressed. Using in situ long-term high-time resolution measurements from NASA's MAVEN mission, we report the first observational characterization of the evolution and parametric dependence of the turbulence energy cascade rate $\varepsilon_C$  at magnetohydrodynamic (MHD) scales. Key findings reveal an averaged three-order-of-magnitude enhancement in $\varepsilon_C$ when transitioning from the solar wind to the magnetosheath. Notably, downstream measurements of oblique and quasi-perpendicular shocks exhibit higher energy dissipation rates than those of quasi-parallel configurations. These results provide the first direct evidence linking shock obliquity to turbulence amplification, offering key insights into shock-mediated turbulence in similar but inaccessible systems. 
\end{abstract}

\newpage
\section*{Plain Language Summary}
Plasma turbulence plays a fundamental role in energy transport within space and planetary environments. However, the injection and dissipation of turbulent energy through the dramatic shock transition are not well understood. Mars, characterized by its compact magnetosheath and extended neutral exosphere, serves as an ideal natural laboratory for investigating solar wind turbulence around a non-magnetized planetary body. Using in situ measurements from NASA’s MAVEN spacecraft, we reveal a distinct evolution pattern of turbulence across the Martian bow shock. Our findings reveal that turbulence energy dissipation is drastically amplified by three orders of magnitude post-shock, with a strong dependence on shock geometry. These results offer novel insights into the evolution of solar wind turbulence in compact, non-magnetized planetary environments, bridging a critical gap in our understanding of collisionless plasma dynamics.

\section*{Key Points}
\begin{itemize}
\item We present the first quantitative evidence of the turbulence dissipation enhancement across the shock transition.
\item Turbulence energy dissipation rate is enhanced by 3 orders of magnitudes on average after shock transition.
\item Energy dissipation rates are higher under quasi-perpendicular shock conditions than under quasi-parallel ones.
\end{itemize}

\section{Introduction}

Turbulence within planetary magnetosheaths is a complex phenomenon characterized by nonlinear energy transfer and fluctuations arising from both upstream transport and local generation processes such as the shock interactions. In the pristine solar wind at 1 AU, magnetic field fluctuation power spectral densities (PSDs) typically exhibit three distinct power-law regimes separated by two spectral breaks. The small-scale break demarcates the transition from magnetohydrodynamic (MHD) scales to ion-kinetic scales \cite{Tu95,Bruno13,Verscharen19,Sahraoui20}. The large-scale break separates the characteristic MHD inertial range scaling ($f^{-5/3}$, where $f$ is the frequency in the spacecraft frame) from a shallower $f^{-1}$ scaling at larger scales. This spectral feature is also commonly observed in planetary magnetosheaths, such as those of Earth, Mars, Venus, and Saturn \cite{Czaykowska01, Alexandrova08, Huang17, Li20, Hadid15, Dwivedi15, Terres21, Ruhunusiri17}. At ion scales, the transition range with steeper power law indices varying from -3 to -6.8 generally emerges between the inertial range and the $f^{-8/3}$ dissipation range \cite{Sahraoui20,huang2021,duan2021,sioulas2023}. However, within the Martian magnetosheath, magnetic field fluctuations frequently display a plateau-like spectral feature, observed in approximately 57\% of {\color{black}\textbf{intervals}}, which exhibits a notable correlation with pickup ion (PUI) parameters \cite{Jiang2023}. Determining the nature of turbulent fluctuations is crucial and novel decomposition methods have been developed to determine the contribution of various MHD modes and coherent structures from the measured parameters \cite{Lion2016, Zank2023}. Previous studies in diverse space plasma environments indicate that the PSD encompasses a mixture of wave modes and coherent structures, including Alfvén waves, fast/slow-mode magnetosonic waves, filamentary Alfvén vortices, current sheets, and magnetic holes \cite{Alexandrova08,Sahraoui2006,Voros2008,Huang17,Lion2016, Jiang2022}. These waves and structures span a wide range of scales and are essential for energy transfer and dissipation \cite{Osman2011,Howes2016,Chasapis2018,Jiang24}.

The magnetosheath plasma undergoes significant transformations due to shock deceleration and compression, leading to pronounced variations in key parameters such as plasma-$\beta$ (the ratio of thermal to magnetic pressure), temperature anisotropy, Alfvénic Mach number, turbulent Mach number, wave characteristics, and intermittency \cite{Schwartz91, Sahraoui2006, Soucek15, Lucek08, Yordanova08, Karimabadi2014, Dimmock14, Li20, Turc23}. Consequently, the properties of turbulent fluctuations in the magnetosheath are strongly modulated by the characteristics of the bow shock and the magnetopause (or, for weakly magnetized planets like Mars, the magnetic pile-up boundary (MPB)). Crucially, both the scale size and magnetic geometry of the bow shock, characterized by the angle $\Theta_{\mathrm{Bn}}$ between the shock normal and the interplanetary magnetic field (IMF), modulate the spectral morphology and nature of downstream turbulent fluctuations \cite{Czaykowska01, Alexandrova08, Li20, Rakh18, Rakh20, jiang_spatial_2025}. During the MHD shock transition, previous studies have suggested that turbulent fluctuation amplitudes increase by one order of magnitude downstream quasi-perpendicular interplanetary shock \cite{Pitna2016,pitna2024,Zank2021}. During a solar coronal mass ejection (CME) event, turbulence has also been examined before, during, and after the CME shock interactions and the highest energy cascade rate was observed in the CME sheath \cite{sorriso-valvo2021,marino_scaling_2023,li_evolution_2017}. However, the evolution of turbulent energy transfer within the Martian magnetosheath and its dependence on specific shock parameters such as $\Theta_{\mathrm{Bn}}$ remain largely unexplored, necessitating comprehensive statistical analyses to resolve these uncertainties.

Mars provides an exceptionally small spatial scale magnetosheath {\color{black}(approximately 5--10\% the size of Earth's magnetosheath), which} imposes fundamental constraints on turbulence development after shock interactions \cite{franco_intermittent_2024}. Moreover, neutral particles that escape from the atmosphere undergo photo-ionization and charge exchange, producing PUIs in the solar wind rest frame with abundant free energy \cite{Cravens87, Chamberlain63, Romanelli16, Rahmati17, Li24}. Injection of energy triggers the development of proton cyclotron waves (PCWs) and other modes within the Martian magnetosheath \cite{Brain02, Delva15, Romanelli13, Harada19, Cowee07, Cowee08, Cowee12}. This localized energy injection complicates our understanding of cross-scale energy transfer in the Martian magnetosheath. Enhanced turbulence energy cascade rates have been observed in the presence of PCWs or mirror modes both upstream and downstream of bow shocks in the solar system \cite{Hadid18, Andres20, Romanelli22}. A one-dimensional hybrid simulation suggests that wave energy in the form of PCWs gradually transfers inversely to larger wavelengths over time, particularly under conditions of strong injection \cite{Cowee08}. The coupling between PUI instabilities and turbulence in Mars' compact magnetosheath—particularly its effects on spectral scaling and energy cascade, poses major challenges for modeling planetary plasma environments.

This study reports the first observational characterization of the turbulent energy cascade across the Martian shock transition, employing exact relations for fully developed compressible turbulence \cite{Andres2017}. Using unprecedented high-resolution measurements from NASA's MAVEN mission, we are able to quantify the turbulent energy cascade dependence on the bow shock geometry and the spatial evolution. Our results demonstrate significant spatial dependence of the turbulence energy cascade, characterized by the most enhanced cascade rates in the nose region of the magnetosheath. Furthermore, we establish functional dependencies of turbulent energy cascade rates on the shock normal angle ($\Theta_{\mathrm{Bn}}$) and the turbulent Mach number.

\section{Data Set and Methods}
\subsection{Data set}\label{subsec41}

We use data from the Solar Wind Ion Analyzer (SWIA) \cite{Halekas15}, the Suprathermal and Thermal Ion Composition (STATIC) analyzer \cite{McFadden15}, and the Magnetometer (MAG) \cite{Connerney15} onboard MAVEN \cite{Jakosky15}. The data set comprises a total of 11,098 magnetosheath {\color{black}intervals} and 6,247 solar wind {\color{black}intervals} from the years 2015 through 2019 \cite{Jiang2023}.

In order to make accurate estimations of the upstream solar wind and the interplanetary magnetic field at Mars, {\color{black}we assume} that the state of the upstream solar wind detected by the MAVEN satellite remains approximately constant within a single orbit period (4.5 hours) and only use the upstream parameters within a single orbit period for each magnetosheath event. First, we eliminate  interference caused by the upstream foreshock. Based on parametric criteria for the unperturbed pristine solar wind used in previous studies \cite{Halekas17}, we identify the pristine solar wind by its velocity $V_\mathrm{i}$, ion temperature $T_\mathrm{i}$, normalized root-mean-square magnetic field perturbation $\sigma_B/B_\mathrm{sw}$, and satellite orbit altitude $L$. We categorize intervals as  pristine solar wind when simultaneously $V_\mathrm{i} > 200$ km/s, $\sqrt{T_i}/V_\mathrm{i} < 0.012$, $\sigma_B/B_\mathrm{sw}< 0.15$, and ${L > 500}$ km, where $\sigma_B=\sqrt{\sum_{i=1}^3 \delta^2 B_i}$ is the root-mean-square value of the magnetic-field perturbation components. 

We estimate the bow shock geometry based on the interplanetary magnetic field data in the upstream pristine solar wind and empirical conic and three-dimensional fitting models for the Martian bow shock (see more details in Supporting Information) \cite{Trotignon06,Gruesbeck2018}. To determine the tangential point, we project the satellite position  along the radial direction to the Mars center onto the bow shock surface \cite{Li20}. Then, we calculate $\Theta_\mathrm{Bn}$ as the angle  between the bow shock normal direction and the interplanetary magnetic field at the projected point. For our statistical analysis, we calculate $\Theta_\mathrm{Bn}$ for each single MAVEN orbit using the corresponding pristine solar-wind parameters.

\subsection{Estimation of the turbulence energy cascade rates}\label{subsec42}

Previous studies {based} their analysis of {the energy cascade rate} in incompressible MHD turbulence {using the so-called Politano and Pouquet relation}  \cite{Politano1998}. Recent studies extend this framework to {isothermal} compressible MHD {turbulence} and successfully apply it to satellite observations and theoretical simulations for estimating  turbulence energy cascade rates \cite{Andres2017,Hadid18,Andres19, Simon2022, B2023, A2023}. 

For adiabatic compressible MHD equations, the compressible turbulent energy cascade rate ($\varepsilon_{c}$) satisfies \cite{Andres2017}
\begin{equation}
-2\varepsilon_{c}=\frac{1}{2} \nabla_l \cdot \mathbf{F_C}+ S_C + S_H + M_\beta,
\label{p6_eq9}
\end{equation}
where $\mathbf{F_C}$ is a flux term proportional to the perturbation two-point increments, $S_C$ is a source term proportional to the divergence of the magnetic field and velocity field, $S_H$ is a combined term of flux and source terms, and $M_\beta$ is a mixed term related to the plasma-$\beta$. Assuming statistically isotropic fluctuations and neglecting all non-flux terms {\cite{Andres19}}, according to Taylor's hypothesis ($l=\mathbf{V} \tau$) \cite{Taylor37}, we obtain 
\begin{equation}
-\frac{4}{3}\varepsilon_{c}l={F_{1C}}+{F_{2C}},
\label{p6_eq10}
\end{equation}
where the flux {terms} ${F_{1C}}$ and ${F_{2C}}$ can be expressed in terms of density $\rho$, velocity $\mathbf{u}$, and Alfvén velocity $\mathbf{u_A}$ {structure functions} as 
\begin{equation}
{F_{1C}}+{F_{2C}}=(\mathbf{F_{1C}}+\mathbf{F_{2C}}) \cdot \hat{\mathbf{V}},
\label{p6_eq11}
\end{equation}

\begin{equation}
\mathbf{F_{1C}}=\langle [\delta(\rho \mathbf{u})\cdot \delta \mathbf{u} + \delta(\rho \mathbf{u_A})\cdot \delta \mathbf{u_a}]\delta \mathbf{u} - 
[\delta(\rho \mathbf{u})\cdot \delta \mathbf{u_A} +\delta \mathbf{u} \cdot \delta(\rho \mathbf{u_A})]\delta \mathbf{u_A} \rangle,
\label{p6_eq12}
\end{equation}
and
\begin{equation}
\mathbf{F_{2C}}=2 \langle \delta\rho \, \delta e \, \delta \mathbf{u}\rangle,
\label{p6_eq13}
\end{equation}
where $e=c_s^2 \ln(\rho/\rho_0)$ signifies the adiabatic closure \cite{Andres2017}. When the density perturbation is zero (i.e., the plasma is incompressible), $\varepsilon_{c}$ in equation \ref{p6_eq10} degenerates to the incompressible MHD turbulence energy cascade rate \cite{Politano1998}. 

\section{Results}
\subsection{Significant amplification of turbulent energy dissipation at bow shock}
\begin{figure*}[!htbp]
\centering
\includegraphics[width=0.95\textwidth]{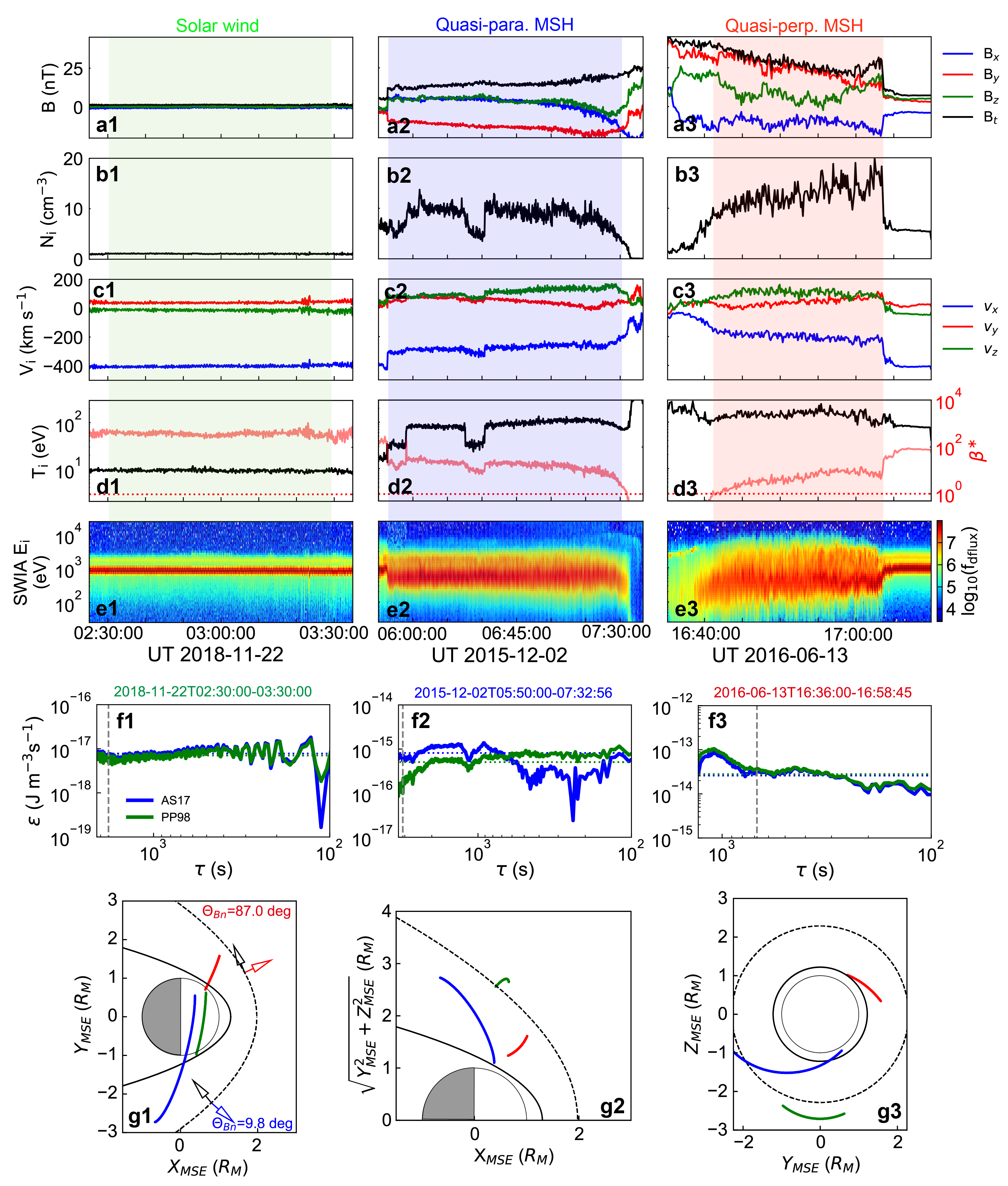}
    \caption{MAVEN observations in the solar wind, quasi-parallel, and quasi-perpendicular magnetosheath regions at Mars. The panels display, from top to bottom, {(a)} the magnetic field {components and magnitude}, {(b)} the ion number density, {(c)} the ion bulk velocity {components}, {(d)} the ion temperature and $\beta^{*}$ {parameter}, {(e)} the ion differential energy flux spectrogram {as a function of time, respectively. The red (blue) shaded areas indicate the magnetosheath regions behind the quasi-perpendicular (quasi-parallel) bow shock. The green shaded areas mark pristine solar wind intervals. Panels (f)} show the {energy} cascade rates as a function of time lag, the horizontal blue and green dotted lines denote the average cascade rates. {Panels (g) display} the {spacecraft's} trajectory in the {Mars Solar Electric} $(X,Y)$ plane, $(X,\sqrt{Y^2+Z^2})$ plane, and $(Y,Z)$ plane{, respectively.} The colored lines in panels {(g)} depict the spacecraft trajectories during the quasi-perpendicular (red), quasi-parallel (blue) magnetosheath and pristine solar wind (green) intervals. The black dashed and solid lines represent the positions of the bow shock and the magnetic pile-up boundary.}
\label{fig1}
\end{figure*}

Figure~\ref{fig1} illustrates results from three distinct segments capturing turbulence characteristics in the upstream solar wind, quasi-parallel magnetosheath, and quasi-perpendicular magnetosheath at Mars, as observed by the MAVEN spacecraft on 22 November 2018, 02 December 2015, and 13 June 2016, {respectively}. The top {four} panels display the magnetic field {components and magnitude}, ion number density ($n_i$), ion velocity {components}, ion temperature {and the} $\beta^{*}$ {parameter, respectively.} Here, $\beta^{*}=2\mu_0(nk_\mathrm{B}T_\mathrm{i}+P_\mathrm{d})/|B|^2$, where $k_\mathrm{B}$ is the Boltzmann constant, $T_\mathrm{i}$ is the ion temperature, and $P_\mathrm{d} = n m_p v_{\mathrm{sw}}^2/2$ is the dynamic pressure. {The fifth panel shows the ion differential energy flux spectrogram. All these measurements are} acquired from the Solar Wind Ion Analyzer \cite{Halekas15} and the Magnetometer \cite{Connerney15} instruments onboard the MAVEN spacecraft \cite{Jakosky15}. The magnetic field and ion velocity are depicted in the Mars Solar Orbital (MSO) coordinate system. The MPB serves as the inner boundary of the magnetosheath and is defined as the location where $\beta^{*}=1$, indicating a balance between plasma (thermal and dynamic) pressure and magnetic pressure \cite{Matsunaga17}. Panels f1 through f3 present the {the incompressible (PP98) and compressible (AS17) dissipation} cascade rates {as a function of the inverse time lag $\tau$} corresponding to the three segments highlighted by green, red, and blue shaded areas in the top five panels. Panels  g1 through g3 show the spacecraft trajectory during the pristine solar wind (from UT 2018-11-22 02:30:00 to 03:30:00), quasi-parallel magnetosheath (from UT 2015-12-02 05:50:00 to 07:32:56), and quasi-perpendicular magnetosheath (from UT 2016-06-13 16:36:00 to 16:58:45) intervals{, respectively.}

We observe distinct characteristics in the {plasma} fluctuations of the solar wind and of the magnetosheath. Following interactions with the bow shock, notable ion heating occurs in the magnetosheath, accompanied by enhancements in magnetic-field strength, ion number density, and velocity fluctuations. To determine the geometries of the bow shock, we assume stationary solar wind conditions and utilize thresholds to estimate the average IMF in the corresponding pristine solar wind intervals for each magnetosheath segment \cite{Halekas17}. Subsequently, we radially project the average spacecraft position onto the bow shock surface defined by the conic model \cite{Trotignon06} and calculate the angle $\Theta_\mathrm{Bn}$  at the projection point. For the quasi-perpendicular (quasi-parallel) magnetosheath, we find $\Theta_\mathrm{Bn}=87^\circ$ ($\Theta_\mathrm{Bn}=9.8^\circ$), as illustrated by Figure~\ref{fig1}g1.

{A}pplying the {exact relation for fully developed turbulence, we estimated the incompressible (PP98) and compressible (AS17)} MHD {energy cascade rates \cite{Politano1998,Andres2017} (see Section \ref{subsec42}). Figure~\ref{fig1} shows the estimation of energy cascade rate both for (f1) the pristine solar wind, (f2) quasi-parallel magnetosheath, and (f3) quasi-perpendicular magnetosheath, respectively.} The gray dashed vertical lines indicate the minimum reliable scale $\tau_0$ (i.e., half of the event duration), above which estimations are unreliable. {\color{black} The magnetic-field spectra in the Martian magnetosheath often present plateau-like triple power-laws \cite{Jiang2023}. This observation suggests that the turbulence in the magnetosheath is not fully developed due to ongoing injection of energy, especially at ion scales. Therefore, we consider the energy cascade rates only for the large-scale inertial range exhibiting a linear scaling in the magnetic-field spectra. We estimate the energy cascade rates by averaging the absolute value of $\epsilon$ over $313~s<\tau<T/2$, where $313~s$ corresponds approximately to the scale where the inertial range ends, as suggested by our statistics of the spectral break frequencies (see more details in the Supporting Information), and $T$ is the total duration of each given interval. For consistency, we calculate all the average energy cascade rates in our following statistical analysis using the same criteria.}

{W}e observe significantly higher magnetosheath cascade rates {when we} compared to those in the {pristine} solar wind. {Specifically, these results do not show a dependence with the bow shock geometry. However, when we compared the quasi-perpendicular {\color{black}interval} with respect to the quasi-parallel {\color{black}interval}, we observe notably higher average cascade rates in the quasi-perpendicular event.} {\color{black}In particular, the incompressible and compressible quasi-perpendicular magnetosheath {cascade rates are $3.35\times10^{-14}$ J m$^{-3}$ s$^{-1}$ (PP98) and $3.11\times10^{-14}$ J m$^{-3}$ s$^{-1}$ (AS17), respectively, while the quasi-parallel magnetosheath cascade rates are  $4.84\times10^{-16}$ J m$^{-3}$ s$^{-1}$ (PP98) and $8.59\times10^{-16}$ J m$^{-3}$ s$^{-1}$ (AS17), respectively.} In comparison, the solar wind turbulence energy cascade rates are $6.78\times10^{-18}$ J m$^{-3}$ s$^{-1}$ (PP98) and $7.59\times10^{-18}$ J m$^{-3}$ s$^{-1}$ (AS17).} {\color{black}We find these values in the solar wind are similar to previous results reported at Mars \cite{Andres20,Romanelli22}, but about one order of magnitude smaller than those reported in previous studies at 1 AU \cite{Hadid17}.} {\color{black}In particular, the turbulent energy cascade rates in the Martian magnetosheath are significantly smaller than those in Earth's magnetosheath \cite{Hadid18,Andres20}.} {We observe an increase of} three orders of magnitude in {the turbulent} energy cascade rates across the bow shock, with average ratios ranging from 3385-4022 for {the} quasi-perpendicular {region} and 72-106 for {the} quasi-parallel {region}. In addition, the quasi-parallel magnetosheath {event} exhibits an enhanced {in the} cascade rate attributed to significant density perturbations, as depicted in Figure~\ref{fig1}f2. The quasi-perpendicular event also shows a significant increase in the density perturbations.

Figure~\ref{fig2} summarizes our statistical analysis of {the turbulent} energy cascade rate in the solar wind and the magnetosheath. {Figure~\ref{fig2}  (a) shows} the compressible MHD turbulent cascade rates estimated for a total of 11,098 magnetosheath {\color{black}intervals} and 6,247 solar wind {\color{black}intervals} spanning from 2015 to 2019 \cite{Jiang2023}. {Figure~\ref{fig2} (b) displays} the histogram of ratios of {turbulent energy cascade rates in the magnetosheath and in the solar wind. The colored bins in blue (green) represent results for compressible (incompressible) turbulent cascade rates. {\color{black}The average compressible turbulent cascade rate is $1.99\times10^{-17}$ J m$^{-3}$ s$^{-1}$ in the solar wind and $2.27\times10^{-14}$ J m$^{-3}$ s$^{-1}$ in the magnetosheath.} {\color{black}The average ratio between the compressible turbulence energy cascade rates in the magnetosheath and the solar wind is very close to that for incompressible cascade rates (1198.44 vs. 1260.11).} {\color{black}In addition, the compressible (incompressible) turbulent energy cascade rates show no significant variation throughout the Martian seasons as a function of the solar longitude of Mars or the level of EUV irradiance, which is directly related to the density of PUIs (see Figure~3 in Supporting Information).}

\begin{figure}[!htbp]
\centering
\includegraphics[width=0.95\textwidth]{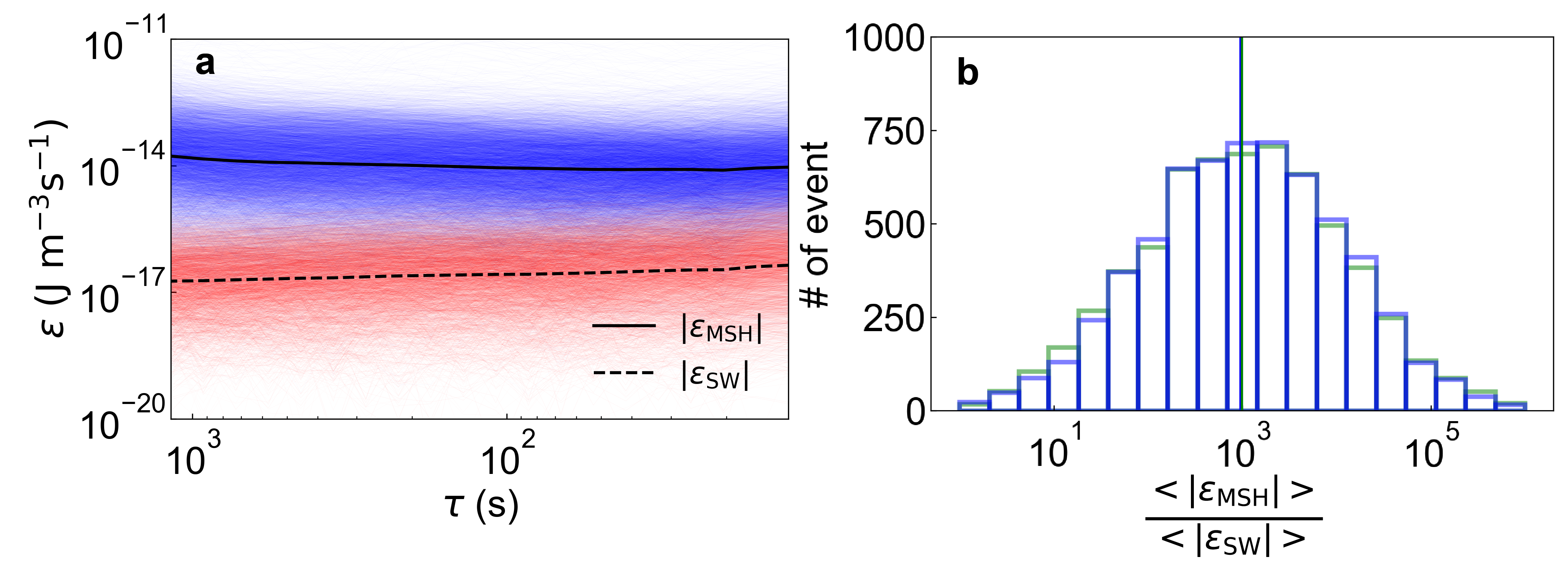}
\caption{(a) Superposition of the compressible turbulent energy cascade rate as a function of time lag for all events. {B}lack solid and dashed lines indicate the average cascade rates. (b) The distribution of the average ratios of compressible (blue) and incompressible (green) turbulent energy cascade rates, {while} the green and blue vertical lines denote {its} average.}
\label{fig2}
\end{figure}

\subsection{Spatial and parametric dependence of the turbulence evolution}

{Figure~\ref{fig3} (a)} shows a {color map} of {the} average compressible cascade rates in the (X, $\mathrm{sign (Y)}\cdot\sqrt{Y^2+Z^2}$) {MSE} plane. Data points are organized into a grid {of} $40\times40$  evenly spaced bins in the ranges $\mathrm{3 > X > -1.7}$ and $\mathrm{3 > \mathrm{sign (Y)}\cdot\sqrt{Y^2+Z^2} > -3}$. The {color represents} average compressible {turbulent energy cascade rate} in each bin. {Figure~\ref{fig3} (b) show} average compressible cascade rate in the magnetosheath under quasi-parallel (red) and quasi-perpendicular (blue) shock conditions as {a} function of {the} distance from the center of Mars. {Figure~\ref{fig3} (c-e)} displays the compressible (AS17) and incompressible (PP98)  cascade rates in the solar wind and magnetosheath, along with the ratio of turbulence energy cascade rates as functions of {the angle} $\Theta_\mathrm{Bn}$. The data {are} binned into an evenly spaced grid with $0<\Theta_\mathrm{Bn} <90^\circ$.

\begin{figure}[!htbp]
\centering
\includegraphics[width=0.95\textwidth]{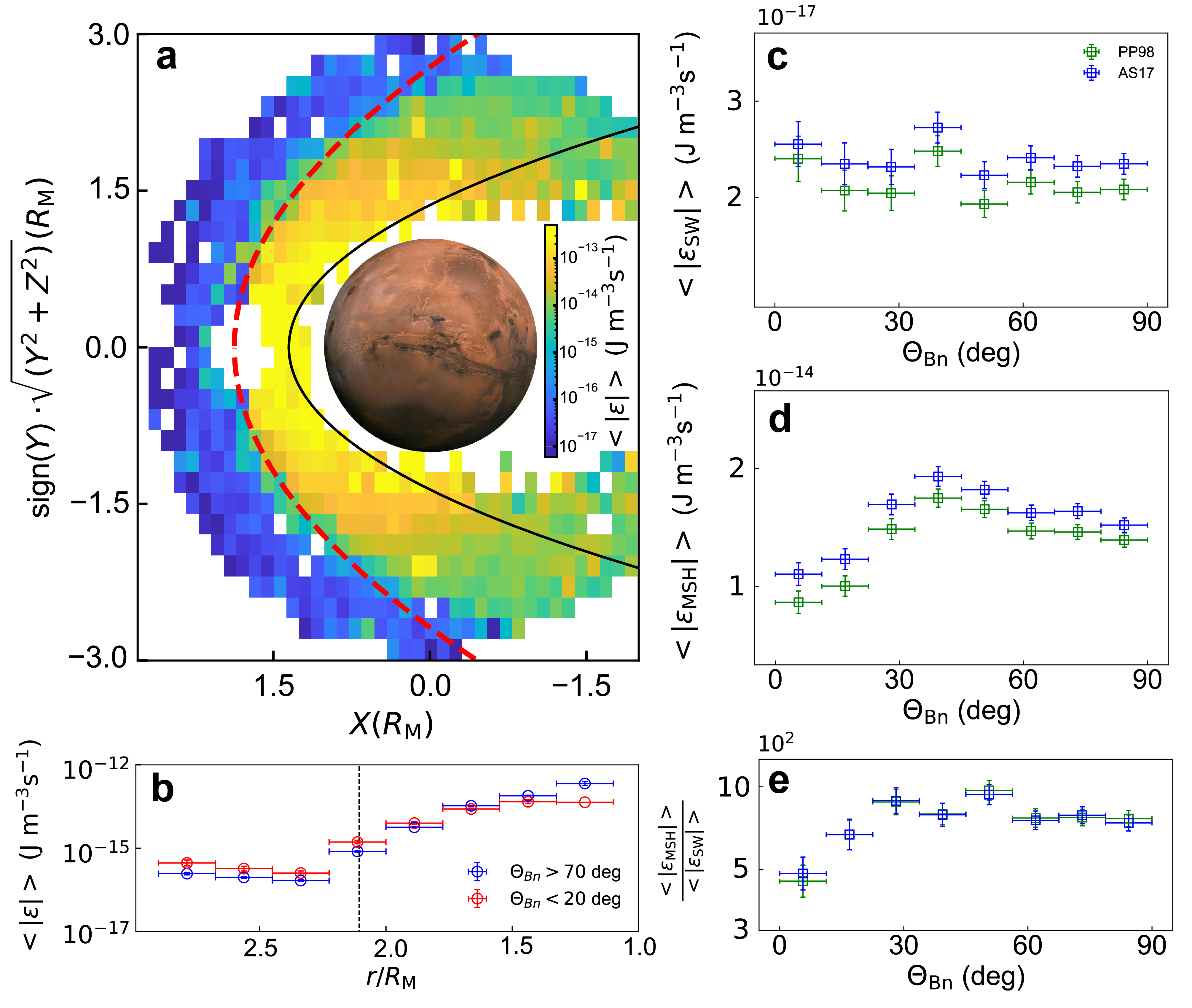}
\caption{(a) Distribution map of compressible turbulence energy cascade rates (AS17) in the $\mathrm{(X,\mathrm{sign (Y)}\cdot\sqrt{Y^2+Z^2})}$ plane in  MSE coordinates. The red dashed and black solid lines denote the nominal positions of the bow shock and the magnetic pile-up boundary. (b) Compressible turbulence energy cascade rates (AS17) plotted against  distance from the center of Mars. The black vertical line indicates the nominal position of the bow shock. Compressible (blue) and incompressible (green) turbulence energy cascade rates as a function of the bow shock normal angle (c) in the solar wind and (d) in the magnetosheath. (e) Ratios of compressible (blue) and incompressible (green) turbulence energy cascade rates from the solar wind to the magnetosheath based on the bow shock normal angle.}
\label{fig3} 
\end{figure}

Our statistical analysis confirms a distinct transition in {the incompressible and compressible} cascade rates when {MAVEN crosses} the bow shock. {More specifically, we observe a significant enhancement in the cascade rate} downstream of the shock, especially in the shock nose region. Radially, the average cascade rates increase approximately by a factor of 1000 when crossing from the solar wind into the magnetosheath. {Figure~\ref{fig3} (c-d)} reveals that compressible cascade rates generally exceed incompressible cascade rates regardless of shock geometry (i.e., independent of $\Theta_\mathrm{Bn}$). In the upstream solar wind, the turbulence energy cascade rates show minimal dependence on $\Theta_\mathrm{Bn}$ and are slightly greater for quasi-parallel bow shock conditions than for quasi-perpendicular conditions. However, in the magnetosheath, turbulence energy cascade rates exhibit a clear dependence on $\Theta_\mathrm{Bn}$ with a peak value around $\Theta_\mathrm{Bn}\approx 45^\circ$. Overall, quasi-perpendicular ($\Theta_\mathrm{Bn} >45^\circ$) magnetosheath turbulence displays greater cascade rates compared to quasi-parallel ($\Theta_\mathrm{Bn} <45^\circ$) magnetosheath turbulence. Figure~\ref{fig3}e shows that the enhancement ratio when crossing the bow shock increases with $\Theta_\mathrm{Bn}$ from approximately 400 ($\Theta_\mathrm{Bn} \approx 10^\circ$) to nearly 1000 ($\Theta_\mathrm{Bn} > 30^\circ$).

\begin{figure}[!htbp]
\centering
\includegraphics[width=0.95\textwidth]{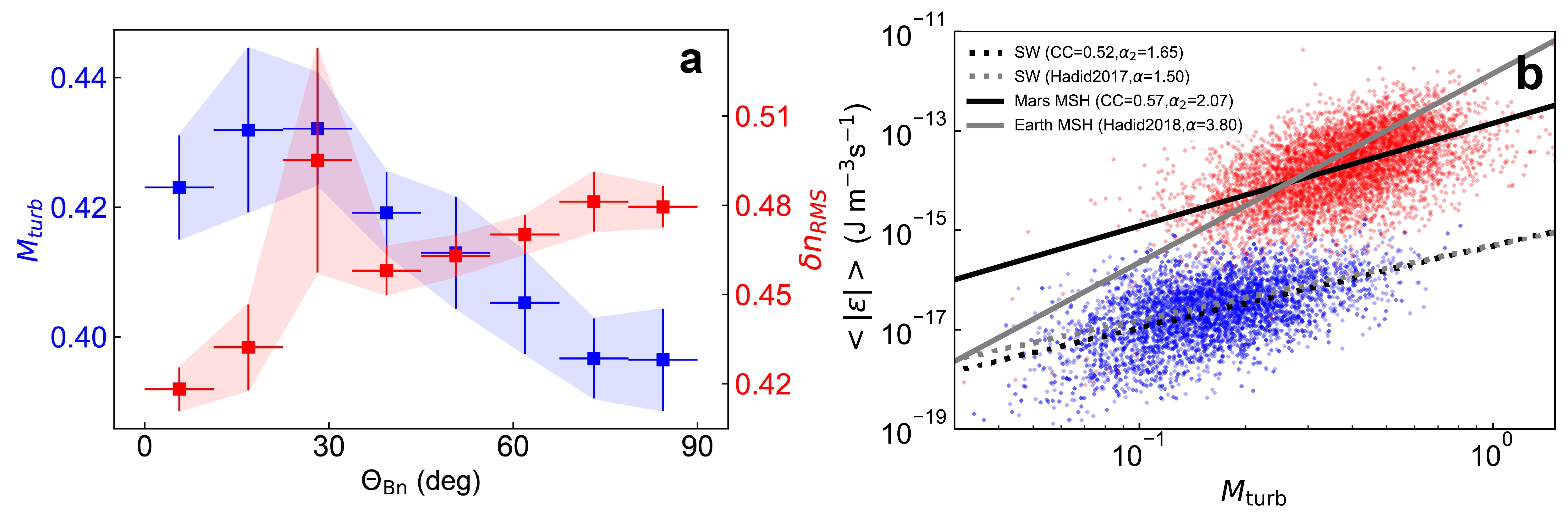}
\caption{(a) The magnetosheath $M_\mathrm{turb}$ and root-mean-square ion density perturbation $\delta n_\mathrm{RMS}$ as functions of the shock normal angle. The error bars represent standard error of the mean value (vertical) and the extent of the bin (horizontal). (b) Solar wind and magnetosheath turbulence energy cascade rates as a function of the local turbulent Mach number $M_\text{turb}$. Dotted and solid lines represent linear scaling laws from fitting and previous studies at 1 au.}
\label{fig4}
\end{figure}

To evidence different levels of density and velocity fluctuations downstream of different shock geometries, we show the local turbulent Mach number $M_\mathrm{turb}=\sqrt{\langle \delta v \rangle^2 /c_s^2}$ ($c_s$ is the sound speed) and the root-mean-square ion density perturbation $\delta n_\mathrm{RMS}$ as functions of the shock normal angle in Figure~\ref{fig4}a. In Figure~\ref{fig4}b, we also find a positive correlation between compressible energy cascade rates in the solar wind and magnetosheath with the turbulent Mach number $M_\mathrm{turb}$, showing correlation coefficients ranging from 0.52 to {\color{black}0.57}. Note that we limit our interpretation of the effect of compressibility to the results of compressible energy cascade rates. Using a linear fitting approach, we find a shallower slope for the energy cascade rate in the Martian magnetosheath (1.93) compared to those previously reported in Earth's magnetosheath (3.8 for Alfvénic events or 4.1 for magnetosonic events, see details in \citeA{Hadid18}). The average $M_\mathrm{turb}$ in the magnetosheath initially increases and then declines with increasing $\Theta_\mathrm{Bn}$, reaching its peak at $\Theta_\mathrm{Bn}\approx 20^\circ$. Unlike $M_\mathrm{turb}$, the density fluctuation level $\delta n_\mathrm{RMS}$ shows a different dependence on $\Theta_\mathrm{Bn}$. Since $M_\mathrm{turb}$ represents a measure of the fluctuating velocity, this result indicates that plasma compressibility could somehow enhance and compensate for the turbulence energy cascade rates in the Martian magnetosheath, leading to a more significant enhancement downstream of quasi-perpendicular shocks. Moreover, the shallow slope for the cascade rates in the Martian magnetosheath suggests that the subsonic ($M_\mathrm{turb}<1$) turbulence is somehow less ``compressible" in a way different from Earth's highly compressible magnetosheath and the nearly incompressible solar wind turbulence. However, there is no theoretical prediction that explains the different scalings between $M_\mathrm{turb}$ and the compressible turbulence energy cascade rates reported by recent in situ observations. Combined with our previous findings on the dependence of the average cascade rates on $\Theta_\mathrm{Bn}$, this result suggests that both $M_\mathrm{turb}$ and $\delta n_\mathrm{RMS}$ are affecting the energy cascade rates in compressible turbulence.

\section{Discussion and Conclusion}\label{sec3}

We present a comprehensive analysis of {compressible turbulence energy cascade rates} upstream and downstream of the bow shock at Mars, {using} high-time-resolution magnetic field and ion data from the MAVEN mission. {To the best of our knowledge, our observational results are the first presentation of a statistical map of {turbulent} cascade in the Martian environment}, revealing a remarkable spatial evolution and location dependence.

Interactions at the bow shock lead to a significant increase in {the turbulent energy} cascade rate. {More specifically,} transitioning from the upstream solar wind to the downstream magnetosheath, compressible turbulent energy cascades experience a substantial enhancement by two {or} three orders of magnitude. In the upstream solar wind, {the energy} cascade rates are evenly distributed in space. In the magnetosheath, however, we observe a significant increase in {the} cascade rates downstream of the bow shock nose, gradually decreasing as the plasma flows towards the flank regions. The cascade rates increase as the spacecraft's distance decreases from the center of Mars, {increasing} by more than a factor of 10 from the bow shock vicinity to the magnetic pile-up boundary. Moreover, our further analysis also suggests that the turbulent energy cascade rates decrease as the magnetic-field spectral index increases. This is in part consistent with a previous study \cite{Andres20}, suggesting that local energy injections from waves generated by PUIs in the magnetosheath may suppress the turbulent cascade. However, there is no significant dependence of the turbulent energy cascade rates on the Martian season or the level of solar extreme ultraviolet irradiance, which is related to PUI processes.

By categorizing the geometries of the bow shock, our results further reveal that the increases in the turbulent energy cascade rates are significantly greater {in the} quasi-perpendicular {region} compared to {the} quasi-parallel {region}. While the solar wind cascade rates remain approximately constant, independently of the shock normal angle ($\Theta_\mathrm{Bn}$), the magnetosheath turbulence energy cascade rates vary, with greater values observed under oblique and quasi-perpendicular shock conditions. In both the solar wind and magnetosheath {plasma}, compressible cascade rates generally exceed incompressible turbulence energy cascade rates, indicating that density fluctuations enhance energy cascade rates as indicated by previous results \cite{Hadid17,Hadid18,Andres19,Andres21,Ferrand22}. We observe a positive correlation between turbulence energy cascade rates in both the solar wind and magnetosheath with the turbulent Mach number $M_\mathrm{turb}$, partially consistent with previous result in Earth's magnetosheath \cite{Hadid18}. Although we find a remarkable agreement of the linear scaling laws in the solar wind turbulence at 1 AU and at Mars, the scaling law in the Martian magnetosheath is quite different from Earth's magnetosheath, requiring further theoretical investigation to explain different scaling laws. Our analysis also suggests that the enhancement of turbulence energy cascade rates in the Martian magnetosheath is influenced by both the levels of density and velocity perturbations. Downstream of quasi-perpendicular shocks, a higher level of plasma compressibility leads to generally higher energy cascade rates compared to quasi-parallel shocks.

Our observational findings offer new insights into the impact of bow-shock interactions on magnetosheath turbulence at Mars. Quantitative analysis shows that the magnetosheath turbulence energy cascade rate is significantly amplified by three orders of magnitude due to the shock compression. Turbulence in the quasi-perpendicular magnetosheath presents a more compressible state compared to the quasi-parallel magnetosheath, hence greater energy cascade rates. However, several challenges persist in understanding the key parameters governing the nonlinear cascade and dissipation of energy. For instance, the exact impact of local instabilities injected by either shock interactions or PUIs in the Martian environment on turbulence energy cascade remains unclear \cite{Ruhunusiri17,Andres20,Romanelli22,Jiang2023,Li24}. {\color{black}We emphasize that our calculations of energy cascade rates cover only the large scales and exclude scales close to the spectral plateau or ion scales. We acknowledge that turbulence may not be fully developed in the Martian magnetosheath, meaning that we cannot strictly guarantee that forcing scales are at the largest scales, dissipation scales at the smallest scale, and a clean inertial range in between. We also cannot be certain that the plasma is homogeneous, which is a strong assumption in the derivation of the third-order laws. Still, we use the third-order laws because they provide accessible information from time series \cite{Huang17,Li20,bandyopadhyay_observation_2021,andres_observation_2023}.} The complex interplay between the nonlinear evolution of waves and background turbulent fluctuations necessitates further investigation. Future multi-point missions, such as the European Space Agency's M-MATISSE science mission candidate \cite{Sanchez-Cano2022}, may offer new insight by simultaneously monitoring upstream and downstream conditions around the Martian bow shock.

\section*{Acknowledgments}
The authors thank the entire MAVEN team for providing the data. This work is supported by the NNSFC grants (Grant No. 42374198, 42188101, 42404177), project of Civil Aerospace ``14th Five Year Plan" Preliminary Research in Space Science (D010302, D010202). W.J. is supported by NSSC Youth grant and the Specialized Research Fund for State Key Laboratories of China. H.L. is also supported by the China-Brazil Joint Laboratory for Space Weather (No. 119GJHZ2024027MI). D.V. is supported by STFC Consolidated Grant ST/W001004/1. {N.A. acknowledges financial support from the following grants: PIP Grant No. 11220200101752 and Redes de Alto Impacto REMATE from Argentina.}

\section*{Competing interests}
There are no competing interests to declare.

\section*{Data Availability Statement}
All data used in the paper and the Supporting Information are publicly available through the Planetary Data System: \url{https://pds-ppi.igpp.ucla.edu/mission/MAVEN/MAVEN/MAG} for MAG \cite{10.17189/1414249} and \url{https://pds-ppi.igpp.ucla.edu/mission/MAVEN/MAVEN/SWIA} for SWIA  \cite{10.17189/1414246}. Data analysis was performed using the Space Physics Environment Data Analysis System (SPEDAS) \cite{2024ascl.soft05001A}.

\end{document}